\documentclass{article}
\pdfoutput=1

\usepackage{arxiv}

\usepackage[utf8]{inputenc} 
\usepackage[T1]{fontenc}    
\usepackage{hyperref}       
\usepackage{url}            
\usepackage{booktabs}       
\usepackage{amsfonts}       
\usepackage{nicefrac}       
\usepackage{microtype}      
\usepackage{graphicx}
\usepackage[round]{natbib}
\usepackage{doi}
\usepackage{float}

\usepackage{booktabs}
\usepackage[normalem]{ulem}
\useunder{\uline}{\ul}{}

\usepackage{array}
\newcolumntype{L}[1]{>{\raggedright\let\newline\\\arraybackslash\hspace{0pt}}m{#1}}
\newcolumntype{C}[1]{>{\centering\let\newline\\\arraybackslash\hspace{0pt}}m{#1}}
\newcolumntype{R}[1]{>{\raggedleft\let\newline\\\arraybackslash\hspace{0pt}}m{#1}}

\title{Why and How Governments Should Monitor AI Development}

\author{ Jess Whittlestone\thanks{Both authors contributed equally.} \\
	Centre for the Study of Existential Risk\\
	University of Cambridge\\
	\texttt{jlw84@cam.ac.uk} \\
	\And
	Jack Clark\footnotemark[1] \\
	Anthropic\\
	\texttt{jack@anthropic.com} \\
}

\date{}

\renewcommand{\headeright}{}
\renewcommand{\undertitle}{}
\renewcommand{\shorttitle}{Why and How Governments Should Monitor AI Development}

\begin{document}
\maketitle


\section*{\textsc{Executive summary}}
We outline a proposal for improving the governance of artificial intelligence (AI) by investing in government capacity to systematically measure and monitor the capabilities and impacts of AI systems.

\textbf{The problem.} Governments face a range of policy challenges from AI technologies, which are developed and deployed at a speed that traditional governance approaches cannot keep pace with. Many of these challenges are caused or exacerbated by the fact that governments lack sufficiently detailed, high-quality, and timely information about AI to govern it effectively.

\textbf{The proposal.} AI as a sector routinely produces a range of data and measures. If these data are synthesized by governments, the insights can improve governments’ ability to understand this technology and its impacts while also helping to create tools to intervene earlier.

Therefore, we propose governments invest in initiatives to measure and monitor various aspects of AI research, deployment, and impacts, including:
\begin{itemize}
    \item Continuously analyzing deployed systems for potential harms, as well as developing better ways to measure the impacts of deployed systems where such measures do not already exist.
    \item Tracking activity, attention, and progress in AI research by using bibliometric analysis, benchmarks and open source data.
    \item Assessing the technical maturity of AI capabilities relevant to specific domains of policy interest.
\end{itemize}

Governments could use this measurement and monitoring infrastructure for a variety of purposes, including:
\begin{itemize}
    \item Testing deployed systems to see if they conform to regulation.
    \item Incentivizing positive applications of AI via measuring and ranking deployed systems.
    \item More rigorous and coordinated approaches to impact assessment and assurance.
    \item Comparative analysis of the strength of countries' AI ecosystems.
    \item Prioritizing funding and incentivizing research.
    \item Early warning systems for sources of risk or opportunity.
\end{itemize}

Governments should play a central role in establishing measurement and monitoring initiatives themselves, while subcontracting out other aspects to third parties (such as through grantmaking, or partnering with research institutions). It's likely that successful versions of this scheme will see a hybrid approach, with core decisions and research directions being set by government actors, then the work being done by a mixture of government and third parties.

Building up this infrastructure will likely need to be an iterative process, beginning with small pilot projects. Promising pilot projects might include:
\begin{itemize}
    \item Assessing the landscape of AI datasets and evaluating who they do and don't represent. Using these findings to fund the creation of datasets to fill in the gaps.
    \item Using geographic bibliometric analysis to understand a country’s competitiveness on key areas of AI research and development.
    \item Hosting competitions to make it easy to measure progress in a certain policy-relevant AI domain, such as competitions to find vulnerabilities in widely-deployed vision systems, or to evaluate the advancing capabilities of smart industrial robots.
    \item Funding projects to improve assessment methods in commercially important areas (e.g. certain types of computer vision, to accelerate progress and commercial application in these areas.
    \item Tracking the deployment of AI systems for particular economically relevant tasks, in order to better track, forecast, and ultimately prepare for the societal impacts of such systems.
    \item Monitoring concrete cases of harm caused by AI systems on a national level, to keep policymakers up to date on the current impacts of AI, as well as potential future impacts caused by research advances
    \item Monitoring the adoption of or spending on AI technology across sectors, to identify the most important sectors to track and govern, as well as generalizable insights about how to leverage AI technology in other sectors.
    \item Monitoring the share of key inputs to AI progress that different actors control (i.e., talent, computational resources and the means to produce them, and the relevant data), to better understand which actors policymakers will need to regulate and where intervention points are.
\end{itemize}

We predict that if such a proposal is not adopted, we will see some version of the following over the coming years:

\begin{itemize}
    \item Private sector interests will exploit the lack of measurement and monitoring infrastructure to deploy AI technology that has negative externalities, and governments will lack the tools available to address them.
    \item Information asymmetries between the government and the private sector will widen, causing deployments to occur that negatively surprise policymakers, which will lead to hurried, imprecise, and uninformed lawmaking. 
    \item Other interests will step in to fill the evolving information gap; most likely, the private sector will fund entities to create measurement and monitoring schemes which align with narrow commercial interests rather than broad, civic interests.
\end{itemize}

\newpage
\tableofcontents
\vspace{4mm}

\section{Introduction}
In this paper we outline a proposal for improving the governance of artificial intelligence (AI) by investing in government capacity to systematically measure and monitor the capabilities and impacts of AI systems. If adopted, this would give governments greater information about the AI ecosystem, equipping them to more effectively direct AI development and deployment in the most societally and economically beneficial directions. It would also create infrastructure that could rapidly identify potential threats or harms that could occur as a consequence of changes in the AI ecosystem, such as the emergence of strategically transformative capabilities, or the deployment of harmful systems.

We begin by outlining the problem which motivates this proposal: in brief, traditional governance approaches struggle to keep pace with the speed of progress in AI. We then present our proposal for addressing this problem: governments must invest in measurement and monitoring infrastructure. We discuss this proposal in detail, outlining what specific things governments could focus on measuring and monitoring, and the kinds of benefits this would generate for policymaking. Finally, we outline some potential pilot projects and some considerations for implementing this in practice.

\section{The problem}
Governments face a range of policy challenges with contemporary technologies, which are developed and deployed at a speed that traditional governance approaches cannot keep pace with \citep{collingridge_social_1980}. This has been especially true of developments in artificial intelligence (AI), where governments have been surprised by technical progress, and lack the tools to intervene in a timely and effective manner to prevent harm. The result is a situation where companies are able to deploy AI systems with substantial potential for harm or misuse in mostly unregulated markets, governments are caught off-guard by these new applications and their impacts, and are unable to effectively scrutinize systems in the ways needed to govern them. 

For example, many governments have recently run into policy challenges relating to:
\begin{itemize}
    \item The rapid deployment and broad diffusion of facial recognition capabilities into the world via chiefly private sector actors (e.g. Clearview AI).
    \item The development of technologies for better editing and manipulation of videos via a cluster of AI techniques colloquially known as `deepfakes'.
    \item Harmful biases displayed by deployed computer vision and natural language processing AI systems.
    \item Radicalization of populations through increasingly effective and opaque recommendation systems deployed on platforms such as YouTube, Facebook, and so on.
\end{itemize}

Many of these challenges can be made more tractable to work on through the addition of more and better information about the underlying aspects of AI technology and diffusion. This is not an unusual suggestion - governments use information and metrics to manage and oversee many critical policy areas. For example, metrics like inflation are critical for managing the economy, data about the prevalence of traffic on major roadways is an input into infrastructure planning, and during COVID-19 we've seen how basic data about the medical status of citizens is a fundamental input into policymaking. However, governments today do not yet use metrics and measures to govern AI in a systematic manner, and the processes governments currently use to get information about AI, such as by convening experts, are too ad-hoc and slow.\footnote{Ad-hoc conventions can sometimes be the best way to gather information, typically for domains where either very little is known, or domains where there have recently been a succession of rapid changes. However, in the context of AI, many things that we might gather ad-hoc groups to analyze will only be easy to identify if we first build continuous monitoring infrastructure to cue us to these developments, or to identify an absence of measurable information.}

It doesn't have to be this way. The policy challenges posed by AI are not inherently illegible: AI as a sector is routinely producing data, metrics, and measures that can be useful for a range of governance purposes. For example, there are well-established metrics for measuring progress in a variety of areas of interest to policymakers, such as: the advance of computer vision capabilities,\footnote{For instance, by analyzing the performance of systems on widely-used datasets such as \textit{ImageNet}.} the emergence of systems for generating `synthetic media',\footnote{Measures such as Frechet Inception Distance (FiD) can help us quantify the quality of synthetic imagery \citep{thakur_how_2021}.} and evaluation suites for understanding the `fairness' of a given AI application.\footnote{There are a broad variety of in-development measures here, ranging from the approach used by `Gender Shades' (covered later in this paper) for computer vision systems, to tools for doing analysis of text-based models.} In more nascent areas of AI research, we often see a proliferation of different measures - there have recently been many proposals for assessing bias in text-generation systems, for example - which can serve as a prompt for governments to convene experts to try and identify consensus measures. The tendency towards open access publishing via preprints on services such as \textit{arXiv} also makes it easy to do large-scale bibliometric analysis of the AI sector as a whole.\footnote{It's also possible to do analysis of peer reviewed papers published via journals, though the costs for accessing this data can be quite high.} 

The policy challenges of AI are also not impossible to anticipate or prepare for: the technologies that are impacting society today have been in research and development for well over a decade. Governments could have been monitoring progress in areas like synthetic media and computer vision and considering their implications well before these capabilities were widely deployed.  There are also promising indications that aspects of future progress are predictable - recent publications on the `scaling laws' of AI systems \citep{kaplan_scaling_2020, henighan_scaling_2020, hestness_deep_2017} give a high-level indication of the capabilities of systems that are yet to be developed. The technologies that most impact society in future will not come from nowhere: they will emerge from the capabilities receiving research attention and seeing progress right now, and we can monitor this attention and progress today. This isn't a novel idea - earlier patterns of progression in technology, such as Moore's Law, have proved to be useful sketches of the future, giving policymakers an ability to plan ahead for certain kinds of predictable technological change (such as, in the case of Moore's Law, the availability of a certain amount of computation for a certain price point by a certain year).

As more AI systems are deployed into the market and technologies advance, they will raise increasingly high-stakes policy challenges \citep{brundage_malicious_2018}, making it increasingly important that governments have the capacity to react quickly to new developments, and the tools to intervene in ways that promote benefits and mitigate risks. These challenges range from those of pervasive and immediate concern, such as the potential for AI systems to reflect harmful biases, to longer-term issues relating to increasingly capable, broad-purpose systems, and the increasing connections between AI capabilities and militarily strategic capabilities. 

\section{The proposal}

In order to keep pace with rapid advances in AI technology, we suggest that \textbf{governments invest in building infrastructure to systematically measure and monitor the capabilities and impacts of AI systems}. Integrating these insights into policymaking would enable governments to produce reliable information about when and where AI is likely to have particularly large societal impacts. This, in turn, would speed up governments’ ability to regulate this technology, while also creating tools to intervene earlier and in ways with a lighter touch than regulation.

We suggest that governments should themselves play a central role in establishing measurement and monitoring initiatives. We expect that developing the necessary infrastructure in-house will have the following benefits:
\begin{itemize}
    \item \textbf{It will be much more likely that the outputs are tailored to and integrated into policymaking}. This is because it will allow governments to play a more direct role in setting the objectives of the new infrastructure, and create a tighter loop between the people doing the measurement and monitoring and other policymakers.
    \item \textbf{It will help develop a set of people in the public sector with the skills needed to analyze increasingly advanced technology}. These people will likely generate unexpected ancillary benefits, such as by advising other parts of government about how to approach areas involving rapidly advancing technology, and helping to form better technical relationships between government and other stakeholders (such as in academia and industry).\footnote{This can also occur via working with quasi-governmental entities - that is, entities with deep links to the government that are integrated into policymaking, such as the National Physical Laboratory in the UK, or entities like MITRE in the USA.}
    \item \textbf{It will lead governments to develop measurement and monitoring infrastructure that can be reused for different purposes}, providing more flexibility around what can be monitored over time, and reducing government dependence on third parties.
\end{itemize}

It may still be useful to subcontract out some aspects of measurement and monitoring to third parties in the private sector or in academia (especially where deeper technical expertise may be needed). However, we suggest that governments need to have a large degree of ownership over and visibility into this work in order for it to strengthen policymaking. In particular, governments should set the objectives for projects and ensure that core infrastructure (e.g., aggregated datasets, search tools, indexes) remains within government (while ensuring it can also be accessible to third parties where needed).

Before getting into the details of what governments should measure and monitor and why, it may be helpful to clarify why we talk here about both \textit{measurement} and \textit{monitoring}, and how they relate to one another.

\textbf{Measurement} means gathering information that reduces our (expected) uncertainty about something \citep{hubbard_how_2014}. The first step for any measurement exercise involves thinking carefully about what needs to be measured, followed by asking whether the requisite measurement tools already exist or need to be created. For example, a government might poll a variety of departments to understand where needs can be met via measurement, and then look for existing tools or work with third parties (e.g. in academia) to build the tools needed.

There are many uncertainties governments might have about AI which more information would help with, for example:
\begin{itemize}
    \item How effective is an AI system at carrying out a particular task?
    \item To what extent does a specific AI system, or type of AI system, possess important societally- or policy-relevant features, such as fairness or robustness?
    \item How are different types of AI capabilities being deployed across different domains of society?\footnote{Note, we imagine governments would regularly analyze what is commercially available, and would not seek to get privileged access to corporate information beyond the information a standard customer could expect to get. We are not advocating that government achieve measurement objectives by gaining privileged access to internal IP.}
    \item What is the rate of growth in deployments for certain types of AI applications?
    \item Which subfields of AI research are seeing the most progress and attention?
    \item What kinds of applications of AI in society might be possible in 2-5 years time, based on what we are seeing today?
    \item How do different countries compare on different aspects of public AI research and development, with a particular emphasis on identifying areas of overlap or specialization with regard to research publishing?
    \item Does a given AI system or company comply with existing regulation?
\end{itemize}

We will suggest that there are measures governments can use to get greater clarity on all of the above questions and more.

\textbf{Monitoring} is the process of operationalizing measurement over time. Once you have ways to measure things you care about, you can establish ways to continually track those measures, enabling insights to be more easily integrated into policymaking, and allowing you to identify meaningful changes.  For some questions, a single measure may be less useful than the ability to monitor changes over time: for example, to identify new capabilities or applications of AI warranting attention, governments need to have a constant stream of data informing them about how these things are evolving.

The combination of measuring and monitoring also creates the ability to perform meta-monitoring of trends in AI over time - that is, understand patterns of development for different areas and how they relate to one another. For instance, if governments were to discover that applications of AI for offensive security purposes were being developed much more rapidly than defensive ones, that might suggest they take actions to improve the balance of development. 

\section{What aspects of AI should governments measure and monitor?}
We suggest two broad categories of things governments could focus on measuring and monitoring: first, the capabilities and impacts of AI systems already deployed in society, and second, the development and deployment of new AI capabilities. We discuss some concrete examples in both these categories below.

\subsection{The capabilities and impacts of deployed systems}

\paragraph{Analyzing systems for potential harms.}
Today, a variety of AI systems are being deployed into the world with potential for great benefits but also great harms, whether that’s as a result of behaving in unintended ways in new environments, being used in harmful ways unintended by their creators, or by displaying harmful biases. Governments could develop tools to analyze and monitor deployed AI systems for robustness, bias, `dual use' capabilities, and other important features.

To illustrate what this could look like in practice, we will talk through an example of analyzing whether a deployed AI system displays harmful biases. By harmful biases, we mean biases that discriminate against the particular attributes of a person in such a way that the system doesn't perform as expected for them. For example, in 2018 the ‘Gender Shades’ project identified ways that commercially deployed computer vision systems from companies IBM, Microsoft, Megvii, Amazon, and Kairos demonstrated inconsistent performance for people of different genders and skin types \citep{buolamwini_gender_2018}. The project was a one-off audit of these systems which led to a broader discussion of bias issues in facial recognition, and IBM directly responded to the project and updated its systems to display less bias.\footnote{Notably, the `Gender Shades' project itself built on a dataset and analytical methodology developed earlier by the National Institute of Standards and Technology (NIST), highlighting how government investment in measurement can unlock subsequent work in academia and other parts of the economy}

Following Gender Shades, in 2019, NIST published the `Demographic Effects' results of its `Face Recognition Vendor Test' \citep{grother_face_2019}. This report saw NIST evaluate more than a hundred distinct facial recognition systems from a range of companies; the organization found significant demographic disparities in performance of computer vision systems, and produced an additional 1200 pages of exhaustive technical analysis, identifying specific issues in different systems.

Both of these approaches exemplify a periodic audit - that is, the audit occurs at some frequency (one off in the case of Gender Shades, while NIST has indicated it may do further studies in the future). Governments could analyze deployed facial recognition systems to generate a continuous stream of information about forms of bias in deployed AI systems.\footnote{Note we're advocating governments test commercially available systems against pre-built datasets, so they can judge performance of these systems, rather than advocating governments test these systems via gathering new potentially privacy-infringing datasets.} This would provide the information policymakers need to identify any systems not complying to known standards, and would also incentivize private sector actors to reduce performance disparities, so as to avoid negative PR or policy repercussions.

There are also a variety of other datasets available that governments may wish to use to evaluate the potential harms of systems, including datasets and tests like \textit{WinoBias} \citep{zhao_gender_2018} and \textit{WinoGender} \citep{rudinger_gender_2018} for analyzing issues of gender representation in natural language processing systems, or \textit{FairFace} \citep{karkkainen_fairface:_2021} for assessing race, gender, and age bias in computer vision facial recognition systems. (There are a multitude of other examples, but providing a full list is out of scope for this paper).

\paragraph{Developing better ways to measure the societal impacts of deployed systems.} The Gender Shades and NIST projects mean there are already established methods to analyze bias in facial recognition systems, making the task of integrating this kind of analysis into policymaking more straightforward. In other cases, governments may try to measure the traits of AI systems related to a given policy concern (e.g. robustness of models or interpretability of outputs), and find there are not already established, usable measures. In such cases, governments themselves can play a valuable role in establishing measures and creating the datasets needed to analyze various societal impacts of AI systems.\footnote{Of course, measures and datasets will not be \textit{sufficient} to assess the wide range of possible societal impacts of AI systems - research focused on understanding how systems interact with social forces will also be needed. However, better measures for assessing societally-relevant features of AI can be a useful starting point for such research: if the robustness or fairness of a system is found particularly low in certain contexts, for example, this can help direct research attention towards better understanding the qualitative impacts in those contexts.}

Governments could fund and coordinate research aimed at developing new approaches to evaluation: for example, ways to evaluate the robustness of different AI capabilities by comparing their accuracy across environments. This could involve funding the creation of various testbeds for evaluating the performance of models. By establishing metrics and evaluation methods for something important like robustness, policymakers could also then use funding or competitions to incentivize further research progress that improves performance on these metrics.

Governments can also help stimulate areas of measurement where there is existing work, but a need to improve tools and establish more continuous measurement. For example, when it comes to assessing fairness in AI systems, there are existing datasets and measures, but not as many as might be needed for fine-grained evaluation of new AI capabilities. In this case, governments may seek to proactively create datasets (rather than use existing ones) that make it easy to test a system for a highly specific form of fairness (e.g. audio speech datasets of different regional accents to test how well speech recognition approaches can serve different accents).

\subsection{The development and deployment of new AI capabilities}

\paragraph{Tracking progress and attention in AI research.}
Many research areas within AI contain widely studied benchmarks and assessment regimes, making it possible to track which subfields of AI research are receiving most attention, as well as where progress on technical metrics is being made within subfields.

For example, analysis by the AI Index on 2020 data showed that robotics and machine learning saw the fastest growth in attention within AI between 2015 and 2020 (based on pre-print publications on \textit{arXiv}), and that computer vision was one of the most popular areas of research within AI in 2020 (31.7\% of all \textit{arXiv} publications on AI in 2020) \citep{zhang_ai_2021}.

Having identified broad areas of interest, we can go deeper and use data on widely studied benchmarks and assessment regimes to analyze which core technical metrics are seeing the most activity within fields. This gives us some indication of the kinds of capabilities that are seeing progress. For example, benchmarks such as \textit{ImageNet} and \textit{SuperGLUE} can be used to monitor progress in computer vision and natural language tasks respectively; and \textit{VoxCaleb}, a dataset and annual competition, provides information about the advancing capabilities of speaker recognition. Monitoring in these areas over the past few years could have alerted governments to the possibility of increased commercial application of these capabilities, which could have prompted earlier investigations into potential sources of bias (by e.g. auditing systems prior to deployment), and other potential areas of societal impact (by e.g. funding earlier research into the impacts and risks of facial recognition). Such awareness could have also let government make more investments into critical areas earlier - such as by prioritizing funding for fairness-related research (as monitoring would have shown a rise in research about the issues of image recognition systems), or proactively creating more diverse datasets to serve as research inputs (as analysis of the papers would have frequently shown a relationship between certain fairness concerns and issues relating to the composition of the input dataset).

We can also directly measure progress on novel capabilities by looking at research papers directly to analyze breakthroughs. For instance, in recent years, generative image models have emerged that can compose synthetic images, and progress can be measured by looking at how well these systems do on the metrics and datasets used by the research community. This approach requires more expertise than judging the winner of an established competition, but may produce the same kinds of insights with significantly fewer resources. This is perhaps the sort of project which governments could define according to policy needs but then outsource to researchers with relevant expertise.

Performance metrics aren't the only way to assess progress in AI, however, and it can also be valuable to understand other factors driving research attention and capability advancement, such as computational costs, data, research networks, and funding \citep{martinez-plumed_accounting_2018}. These items serve as some of the key inputs into the development of AI technology, so monitoring them can give governments a sense of trends that may influence downstream development, or serve as useful signals in their own right.\footnote{For instance, a dramatic rise in the computational intensity of one type of research would be useful for funding bodies to be aware of, in order to create new funding streams for promising technical areas. Similarly, a drastic reduction in the cost (and/or increase in availability) of useful datasets might indicate imminent commercialization and deployment of certain AI technologies.} These inputs may sometimes be able to provide earlier `warning signs' of important progress than performance metrics or monitoring deployed systems would, as we’ll discuss later.

For example, increases in the size or complexity of datasets used for training models can be a valuable indicator of progress in a field. In computer vision, for instance, we've seen companies move from training on millions of images to billions, and there has also been a shift to more complex datasets (e.g. from datasets with ten different types of `vehicles' to hundreds of different types). Closer analysis of these trends could help identify ways that computer vision systems are becoming more capable and/or robust across different domains.

Similarly, we can analyze compute and its relationship to progress in AI by reading papers and assessing the amount of computation that has been used to achieve a given AI capability \citep{amodei_ai_2018}. By doing this form of measurement, governments can identify the resource intensiveness of certain types of research, and develop intuitions about which researchers are capable of using large-amounts of computation in their experiments, leading to insights about where progress in an area is variant or invariant to the usage of computational resources. Likewise, governments could roughly estimate the amounts of money being used to train certain large-compute systems, which would help provide better information for shaping funding of grants to researchers - if the cost of compute for certain projects increases by ten to a hundred times, government should ensure it can fund some researchers to do experiments at this scale. Such insights can subsequently be used to develop science funding strategies as well as to help governments model out the global competitive landscape with regard to compute and AI research. All of this combined could help a government identify and develop strategic AI capabilities.

\paragraph{Assessing the maturity of capabilities in specific domains.}
Governments could also pick specific domains which they see as high priority, and focus on measuring and monitoring the state of AI research and deployment in those domains. 

This would make it easier to prioritize investments in areas of AI research that most directly serve policy priorities, such as developing commercial ecosystems or supporting national security interests. Governments could systematically monitor areas of interest to develop high-quality data about given technical capabilities (and research fields); such information would have a range of downstream uses, from simply flagging people with expertise, to alerting governments to areas where they may want to further fund scientific investigation or parallel lines of research that support or counteract an emerging capability. Below we outline two examples of this in a commercial and a national security context.

\textit{Commercial:} There may be areas of AI development that hold particular commercial relevance for a given government. By monitoring the state of AI development in these domains, governments can assess how competitive they are at given AI capabilities, and analyze if they have national assets that they can bring to bear to further stimulate activity in these areas.

For instance, in the case of the UK, the National Health Service is a potential source of valuable data which could be used to develop useful technologies for applying AI to healthcare. The UK has also set up NHSx to help improve the way the NHS uses digital technologies. NHSx could, in partnership with other government stakeholders, evaluate the state-of-the-art for AI as applied to medical diagnostics, identify areas where the NHS (or NHSx) can create or unlock data which could help further develop these capabilities, then seek to stimulate commercial activity here by hosting competitions to improve capabilities in specific areas.\footnote{Of course, care should be taken to ensure this analysis respects important privacy concerns.}

\textit{Security:} There are areas of AI development that hold relevance for national security, either via enabling new capabilities, creating new defenses against existing capabilities, or radically reducing the cost of carrying out certain security-relevant functions (such as surveillance). All of these are critical areas for governments to have high awareness of, and better measurement and monitoring provides a way to identify emerging security concerns or changes in the landscape.

For example, it might be useful for governments to have a more detailed picture of the state of security-relevant AI capabilities that are visible in public, open source research and code. These capabilities could include areas like pedestrian re-identification (an AI capability that uses computer vision to track and analyze pedestrians as they traverse an area, then re-identify them when they appear in another camera feed), or speaker recognition (a capability which lets you recognize a specific speaker in audio data), or the development of AI systems that are able to autonomously defend or attack other software systems. By analyzing this data, governments could identify technical trends driving progress in security-relevant capabilities, better understand which institutions (e.g. universities or companies) are contributing to the development of these areas, and map out the international research networks that exist for a given area of interest. All of this data can better orient governments to the future of security as augmented by AI, and the synthesis of such data may yield insights that can give a government a strategic edge - for instance, noticing a spike in publishing and capability advancement on AI for cyberoffense may prompt governments to increase funding in cyberdefence. Additionally, some aspects of security-related monitoring will naturally rely on private (classified) data only available to government actors, or could involve the private sharing of confidential information between industry and government about emerging threats.

\paragraph{Developing better ways to assess progress.}
Some areas of research have more established progress metrics than others, and some will be more directly relevant to policymaking than others. Governments might therefore also identify areas or types of progress that are important for policy priorities (such as those tied to deployments in specific domains, or those with potentially large societal impacts), and seek to establish more rigorous ways to measure and monitor them.

For example, for areas like computer vision, it would be valuable for governments to be able to not just monitor progress in underlying techniques, but to understand how the safety and robustness of these systems are evolving relative to the underlying techniques, and which actors are contributing to progress in different areas. Policymakers could build on existing datasets such as the \textit{ImageNet} dataset to evaluate and monitor which research groups are capable of further developing the state-of-the-art, helping to make governments aware of useful clusters of expertise.

Governments could also combine different sources of data into useful composite measures relatively easily. As image recognition systems have matured, researchers have created variants of \textit{ImageNet} to help them better assess progress, including \textit{ImageNet-A} (short for ImageNet Adversarial), a collection of so-called `natural adversarial examples' - images that are inherently challenging for contemporary AI systems to label correctly, and \textit{ImageNet-R} (short for ImageNet Rendition): a collection of stylized versions of some of the things depicted in the original \textit{ImageNet} dataset (for instance, cartoons or sculptures of bears, versus photographs of bears in the original dataset). These can be used to assess how robust computer vision systems are and how well they generalize. By tracking both progress on  \textit{ImageNet} \textit{and} progress on harder \textit{ImageNet} variants, governments could measure how progress in computer vision capabilities compares to progress in robustness and generalizability, highlighting areas of risk and informing funding decisions.

More ambitiously, governments could incentive the creation of entirely new measures in areas that are currently hard to evaluate.
 Robotics is a good example. Progress in combining contemporary AI methods with robotics could bring many benefits, such as reducing costs in industrial manufacturing, increasing the capabilities of drone platforms and creating new capabilities for responding to natural disasters. Such progress could also change the strategic balance of military power between nations, and change the susceptibility of many jobs to automation, making it an important area for policymakers to monitor. From 2015 to 2017, Amazon held an annual competition to improve the state of development in robotics, by evaluating how well robots could do at picking and stowing items on shelves. They held three competitions over three years, with each competition generating information about how well contemporary robots and AI algorithms could be adapted to an industry-relevant challenge. Amazon retired the competition after 2017 - the understanding of the authors is that this was because progress had not been as rapid as Amazon had hoped.

Better evaluation could help speed up progress, but
robotics is difficult to evaluate for a few reasons. Physical reality is highly variable, making direct comparisons between capabilities developed by different labs challenging. Additionally, the cost of conducting robotic experiments is typically high, as it requires access to a physical testing environment, as well as support for the hardware (which will, inevitably, break in a few ways during testing). Given the policy relevance of progress in robotics, governments may want to consider how to create better approaches to robotic testing, which could both be used to incentivize research and to monitor signs of progress.

\section{How could governments use AI measurement and monitoring?}

In general, the proposed measurement and monitoring infrastructure aims to both: (a) provide governments with the \textbf{information} needed to make better decisions around AI; and (b) provide governments with the ability to exert greater \textbf{influence} over the AI ecosystem, by creating incentives to positively shape AI development and deployment. 

Governments could use the information generated by this infrastructure to perform a variety of tasks, including: identifying areas where growth could be catalyzed by funding research; identifying potential benefits and harms of deployed technologies; and building early warning systems for areas undergoing rapid and consequential technological change.

Governments can also use the information generated by measurement and monitoring exercises to exert greater influence over the AI ecosystem - as the old adage goes, information is power. Governments can use better information to play a greater role in discussions about what things in AI should be measured, how they should be measured, and what should be prioritized and when. Similarly, gathering information about the state of deployed AI systems, their capabilities, and where they're being deployed, can give governments a greater ability to identify areas where it may wish to support further deployments in the economy, or areas where it may want to take a more active regulatory role.

\subsection{Testing deployed systems to see if they conform to regulation}
If governments can monitor various features of AI systems, such as their susceptibility to different types of bias and their robustness across a variety of environments, then it will be much easier to identify whether they conform to existing regulation (such as the EU’s new proposed AI regulation).\footnote{Though in most cases it won’t be the role of governments to actually assess whether systems conform to the EU regulation, we do think it is important that governments play a key role in establishing what those assessments should involve, to avoid this being captured by private sector interests. Better measurement and monitoring would make it easier to set clear standards for this kind of assessment.} Carrying out this kind of measurement and monitoring exercise will also create knowledgeable staff within government who can design new governance approaches based on a rigorous understanding of AI capabilities and impacts.

AI systems are not static products - they are frequently updated with new datasets, as well as tweaks made possible by the data generated via interaction with their users. For instance, a computer vision system for analyzing satellite imagery will be continuously updated with new satellite imagery as it is developed, or a facial recognition system may be updated with additional photos of people already in the system, or new pictures of previously underrepresented groups. Therefore, when governments craft regulations for AI systems (e.g. ensuring a given system is conforming to a pre-agreed level of bias according to some metric), they will need to be able to check that deployed systems continue to conform to the requirements, which calls for continual monitoring.

\subsection{Incentivizing positive applications of AI via measuring and ranking deployed systems}
Once a government has begun to measure and monitor a salient policy trait, they can synthesize this information and publish indexes that rank systems and research via these traits. For instance, there are techniques that researchers use to assess the fairness (or lack of fairness) of deployed AI systems which governments could use rank the `most fair' systems. This would have a range of potentially positive effects. First, it would give technology consumers and buyers better information about products; second, it would incentivize companies to compete with each other to field systems that could be at the top of such rankings; and third, it would be a disincentive for companies to deploy harmful systems (as they'd likely be measured and ranked, bringing adverse policy and public relations issues).

Similarly, we could imagine exercises in measuring the computational efficiency of certain AI systems, letting governments rank products according to their environmental impact. This could create the data to help build an AI equivalent of the \textit{Energy Star} energy efficiency ranking program, or similar.

\subsection{More rigorous and coordinated approaches to impact assessment and assurance}
Despite plenty of research and policy-focused work, there’s still no accepted or standardized approach to impact assessment or assurance of AI systems, which will be crucial for mitigating risks and enabling benefits. Several different groups are calling for better coordination in establishing assessment and assurance methods \citep{project_sherpa_recommendation:_2020, ahamat_need_2021, committee_on_standards_in_public_life_artificial_2020}. The new EU regulation - and its requirements that high-risk systems undergo conformity assessments - makes the need for standardization and coordination on impact assessment more pressing. If the regulation is going to be effective, it needs to set clear baseline standards for what conformity assessments should involve. Without active government involvement these standards risk being captured by industry, allowing them to `mark their own work'.

Better defined measures for aspects of responsible AI - including safety, fairness, and robustness - will make it much easier to establish clear processes for impact assessment. If governments and regulators can play an active role in the development of these measures, they are more likely to be developed in ways that serve the public interest.

\subsection{Comparative analysis of the strength of countries' AI ecosystems}
We could use this measurement and monitoring infrastructure - continually collecting information on the features of deployed AI systems, and on progress in AI research and applications - to help us compare and contrast different countries' AI ecosystems and research capabilities. This can help orient policymakers as to where their country’s particular strengths and weaknesses are.

Today, various policy entities are trying to construct indexes which can be used to evaluate the domestic traits of AI capabilities, research communities, and ecosystems across different nations. This can take the form of specific reports looking at certain technical capabilities or research publishing areas,\footnote{For instance: work from the Center for Security and Emerging Technology \citep[e.g.][]{toney_comparing_2021}.} or attempts to holistically compare nations according to their AI performance on a range of distinct areas.\footnote{For instance: work from the OECD's AI Policy Observatory \citep[e.g.][]{oecd.ai_database_2021}, and work from the AI Index at Stanford University \citep[e.g.][]{zhang_ai_2021}.} However, such approaches tend to be hosted at institutions that lack government-level access to country data (or the ability to incentivize local actors to produce such data), and sometimes at institutions that have the potential to be relatively short-lived (for instance, projects tied to specific private think tanks or companies have less of a guaranteed long-term focus than mandated government functions).

If policymakers are able to gain an accurate picture of where their AI ecosystem stands with regard to other countries, then they can more confidently carry out a range of activities that can benefit their nation, as well as aid international coordination. These include:
\begin{itemize}
    \item Identifying areas of research strength to maintain leadership in.
    \item Identifying areas of weakness; discussing with their research base if this weakness matters for the country, and if it does, adapting funding to improve things here.
    \item Identifying `research clusters' that span between their country and other countries, which would reveal opportunities for funding and diplomatic collaboration around common technical goals.
\end{itemize}

\subsection{Prioritizing funding and incentivizing research}
Policymakers today often lack well calibrated intuitions about their country’s AI capabilities and capacity, which makes it difficult for them to identify specific priorities for the broader research community. 

Measurement and monitoring infrastructure can both help target research funding in the most effective areas, and provide policymakers with more effective tools to incentivize research. By combining bibliometric analysis of how much attention a research area is receiving, with analysis of widely studied benchmarks to see how much progress that area is seeing, governments can identify areas where funding may be most needed. This includes identifying areas where there are good opportunities for increased funding (if a field is seeing a lot of progress on benchmarks but possibly receiving comparably little attention), as well as where the research community are facing challenges that need more targeted funding to overcome (if there is a large amount of publishing in an area but little progress on metrics). In addition, if governments can robustly measure the things they care about, they can more easily create incentives for research and industry to build systems that perform better on these measures, through funding or competitions. For example, better measures of different aspects of `trustworthy AI' - fairness, interpretability, robustness etc. - could allow governments to identify the places where research currently falls particularly short, and more effectively target funding to ensure the AI systems deployed in society can gain public trust.

By investing in measurement and monitoring, policymakers will be better equipped to identify areas where more research can support a policy need - and that research, in turn, is likely to generate more useful information for policymakers. For example, a better understanding of the metrics currently used in research to assess the fairness of AI systems would enable policymakers to identify specific types of fairness that aren't being evaluated, and push for more work to reduce these gaps. 

\subsection{Early warning systems for sources of risk or opportunity}
If governments are collecting and analyzing data on various dimensions of AI research and deployment, they can monitor for signs of progress likely to impact society, and be less likely to be caught off guard by new developments. 

For example, for some types of AI system - such as language models - there is a direct relationship between the amount of computational power used in their development and their potential for downstream beneficial and harmful impact \citep{kaplan_scaling_2020}. As many of these systems are developed by actors in the private sector, monitoring where large amounts of compute are being used may enable governments to develop early awareness of which actors are likely to be developing and deploying highly capable systems.

Combining monitoring of progress in AI capabilities and applications with measures of their potential societal impacts could also be used to identify areas where more governance is needed. If many systems are being deployed that perform poorly on important measures such as fairness or responsibility, this highlights an area for potential policy intervention. More broadly, by combining monitoring of progress in fundamental AI capabilities with more rigorous analysis of their possible societal impacts, it may be possible to identify potential applications where governance needs to be thought through ahead of time. For example, if we begin seeing progress towards being able to deploy more autonomous AI systems in the world (such as progress in the real-world applicability of reinforcement learning), this could be an important prompt for developing new approaches to oversight \citep{whittlestone_societal_2021}.

\section{Implementation}

For this measurement and monitoring infrastructure to have the desired impacts, it needs to be effectively integrated into policymaking. Doing this in practice is not necessarily straightforward: a natural failure mode for this proposal would be for a government to invest in a `measurement body' that doesn't quite connect to the right parts of government and doesn't quite produce the right kinds of information to feed into specific policy decisions. Key to avoiding this failure mode will be to align the initial projects done by the measurement and monitoring team(s) with clear policy-led demands.

It may therefore be sensible to start with some initial pilot projects that address a clear policy need with clearly identified government stakeholders - by conducting these pilots, we can better understand the process by which insights are generated and shared across government. Starting with these small pilot projects can also help build up the in-house expertise and infrastructure needed to conduct more ambitious projects, as well as providing a proof of concept, making it easier to generate more funding and support. With this in mind, some examples of the kinds of projects that could work well at this pilot stage, based on some of the earlier selected case studies, include:

\begin{table}[H]
\begin{tabular}{ L{0.5\textwidth}  L{0.5\textwidth} }
\toprule
\textbf{Policy challenge} & \textbf{Pilot project} \\ \midrule

Identifying and mitigating harmful biases in machine learning models. &
  Assessing the landscape of AI datasets and evaluating who they do and don't represent. Using these findings to fund the creation of datasets to fill in the gaps. \\ \midrule
  
Understanding how competitive a nation's AI ecosystem is and where there are opportunities to improve. &
  Using bibliometric and technical analysis to understand areas in which a country is doing well, and where it may be falling behind. Using these insights to prioritize government investment, or to identify opportunities to coordinate with other countries. \\ \midrule
  
Understanding progress in policy-relevant AI domains (e.g vulnerabilities in widely-deployed computer vision systems, or the advancing capabilities of smart industrial robots). & Hosting competitions to make it easy to measure progress in such domains. \\ \midrule
  
Accelerating progress and commercial application. &
  Stimulating activity in commercially important areas (e.g. certain types of computer vision) by funding projects to improve assessment methods. \\ \midrule
  
Understanding how advances in AI may alter the types of jobs people do. & 
   Tracking the deployment of AI systems for particular economically relevant tasks, in order to better track, forecast, and ultimately prepare for the societal impacts of such systems. \\ \midrule

Ensuring policymakers are aware of the potential harms of AI systems that are being developed and deployed. &
   Monitoring concrete cases of harm caused by AI systems on a national level, to keep policymakers up to date on the current impacts of AI, as well as potential future impacts caused by research advances - similar to the `AI Incidents Database' initiative launched by the  \citet{partnership_on_ai_artificial_2020}. \\ \midrule

Understanding AI deployment on a sector-by-sector basis. 
& Monitoring the adoption of AI technology across sectors. One measure for this could be counts of deployed systems in given sectors. \\ \midrule

Knowing which actors policymakers will need to regulate and where intervention points are.
& Monitoring the share of key inputs to AI progress that different actors control (i.e., talent, computational resources and the means to produce them, and the relevant data). \\ \bottomrule
\end{tabular}
\end{table}

Depending on government resources and existing bodies, measurement and monitoring infrastructure could be implemented on different levels and with different kinds of partnerships with third party organizations. A minimal version of this proposal would be to identify teams within existing agencies who already do similar things, where funding could support additional projects; more ambitiously it might be possible to establish an entirely new team or even a new department to conduct this work. 

Mixed models, spearheaded by the government but involving significant contributions from outside organizations, could also be effective: one option would be a network drawing on expertise from different parts of government and academia with a central team coordinating and synthesizing. Since the details of this will heavily depend on the institutional context, we intend to spend significant time discussing this proposal and the specifics of its implementation with a range of policymakers before making more detailed recommendations.

\section{Conclusion}
This paper outlines an approach to the measurement and monitoring of AI systems which, if adopted, would better equip governments to respond to the policy challenges posed by the development and deployment of increasingly powerful AI systems.

We predict that if such a proposal is \textit{not} adopted, we will see some version of the following over the coming years:
\begin{itemize}
    \item Private sector interests will exploit the lack of measurement and monitoring infrastructure to deploy AI technology that has negative externalities, and governments will lack the tools available to address them.
    \item Information asymmetries between the government and the private sector will widen, causing deployments to occur that negatively surprise policymakers, which will lead to hurried, imprecise, and uninformed lawmaking. 
    \item Other interests will step in to fill the evolving information gap; most likely, the private sector will fund entities to create measurement and monitoring schemes which align with narrow commercial interests rather than broad, civic interests.
\end{itemize}

Instead, if governments adopt measurement and monitoring schemes, we predict they'll rapidly convert this investment into direct and indirect benefits. The direct benefits will include the provision of better information for funding agencies, a greater degree of `legibility' of the strengths and weaknesses of domestic researchers, early indications of areas of technical change that may have policy implications, and systems for detecting compliance with regulatory regimes. The indirect benefits may be even greater - namely, creating the expertise and infrastructure within government to analyze and understand a rapidly changing technology landscape. We will be seeking to work with governments and other interested stakeholders to further develop and operationalize these proposals, and plan to pursue specific case studies for future research.

\section*{Acknowledgments}
We thank Sam Clarke, Luke Muelhauser, Dario Amodei, Ghazi Ahamat, Shahar Avin, Tabitha Goldstaub, Danny Hernandez, Angus Mercer, Iason Gabriel, Don Wallace, Alexandra Belias, Lewis Ho and Matt Clifford for helpful discussions and feedback on drafts of this work. We are especially grateful to Sam Clarke for significant feedback and editorial support.

\bibliographystyle{unsrtnat}
\bibliography{citations}

\end{document}